\documentclass[11pt]{article}
\usepackage{geometry}
\usepackage{cite}
\usepackage{amsmath}
\usepackage{amssymb}
\usepackage{amsfonts}
\usepackage{graphicx}
\usepackage{caption}
\usepackage{subcaption}
\usepackage{dcolumn}
\usepackage{float}
\usepackage[usenames,dvipsnames]{xcolor}
\usepackage{tikz}
\usepackage{pstricks}
\usepackage{lipsum}
\usepackage{appendix}
\usepackage{authblk}
\usepackage{verbatim}
\usepackage{setspace}
\usepackage{gensymb}
\usepackage{tikz}
\usepackage{hyperref}
\hypersetup{
	colorlinks=true,
	linkcolor=blue,
	filecolor=cyan,      
	urlcolor=magenta,
}
\begin{document}
\title{Deviation in stellar trajectory induced by asymmetry in partial tidal disruption} 
\date{}
\author{Pritam Banerjee \thanks{bpritam@iitk.ac.in}}
\author{Debojyoti Garain \thanks{dgarain@iitk.ac.in}}
\author{Shaswata Chowdhury \thanks{shaswata@iitk.ac.in}}
\author{Dhananjay Singh \thanks{sdhanjay@iitk.ac.in}}
\author{Rohan Joshi \thanks{rohan@iitk.ac.in}}
\author{Tapobrata Sarkar \thanks{tapo@iitk.ac.in}}
\affil{Department of Physics, \\ Indian Institute of Technology Kanpur, \\ Kanpur 208016, India}
\maketitle
\begin{abstract}
We study partial tidal disruption and present a quantitative analysis of the orbital dynamics of the remnant self-bound core. We perform smoothed particle hydrodynamical simulations to show that partial disruption of a star due to the tidal field of a black hole leads to a jump in the specific orbital energy and angular momentum of the core. It directly leads to deviation in the core's trajectory apart from getting a boost in its velocity. Our analysis shows that the variations in the specific orbital energy and angular momentum are higher when the pericentre distance is lower. We conclude that higher mass asymmetry of the two tidal tails increases the magnitude of the trajectory deviations. Our study reveals that observable deviations are only possible when mass ratio $q \lesssim 10^3 $, which indicates the range of intermediate-mass black holes. 

\end{abstract}
\section{Introduction}

Stellar objects are often tidally deformed and disrupted in the vicinity of supermassive black holes (SMBH) found near the galactic centers \cite{Lacy, Rees}. Similar events can also occur with smaller black holes scattered around galaxies and globular clusters. When a star is ripped apart by the strong tidal forces of a black hole, tidal debris falls back to the black hole and forms an accretion disk \cite{Evans, Kochanek, Guillochon2014}. In the process, shock-heated material emits electromagnetic flares that can be observed when it is above the Eddington luminosity \cite{Komossa, Gezari, Burrows, Cenko, Bogdanovic}. Tidal disruption events (TDE) are an important source of information to understand the interior properties of stellar objects, nature of the black hole, as well as stellar population and dynamics in the vicinity of it \cite{Lodato, Bonnerot}. The luminous flares emitted from a TDE is a signatory evidence of a black hole that is otherwise quiescent. It also provides a powerful diagnostic of the internal properties of the disrupted star and the composition of the debris \cite{Guillochon2013, Rosswog2009, Kochanek2016a, Kochanek_etal2016, Mockler2022}. Tidal events are vastly studied in the literature and there is a fair amount of theoretical understanding of such astrophysical phenomena. However, with the advent of powerful numerical simulations, TDEs are frequently revisited in the literature to decipher new and important aspects of tidal interactions, which are crucial for understanding relativistic astrophysics in the vicinity of a black hole.

One such important aspect was introduced by Manukian et al. (2013) in \cite{Manukian}, who demonstrated that a stellar object, after being partially disrupted by a black hole, recoils back to form a core due to its self-gravity. During this process, it gains a `kick' in its orbital velocity owing to the boost in the specific orbital energy of the remnant bound core. The asymmetric mass distribution was shown to be the cause of the phenomenon. The asymmetry in the tidal tails exerts a net force on the self-bound core due to the conservation of linear momentum and turns it into a turbo-velocity star (see \cite{Brown}) that can exceed the escape velocity at the stellar surface. Rosswog et al. (2015) \cite{Rosswog_tidalkick} extend this study incorporating relativistic effects in the tidal encounter of a solar-type star due to an SMBH. They show that relativistic encounters are significant in enhancing the kick velocity of the core. It is well known that an orbit in spherically symmetric static space-time is defined by two constants of motion: specific orbital energy and specific orbital angular momentum (OAM). In the previous works, a lot of importance is given to the boost of specific orbital energy of the remnant core due to partial tidal disruption. However, as per our knowledge, there is a lack of quantitative study in the literature on the change of specific OAM of the remnant core. In this paper, we undertake this task. Any change in specific OAM changes the trajectory of a particle. Therefore, similar to a gain in velocity due to a change in orbital energy, trajectory of the remnant core is expected to change in the aftermath of a tidal interaction with a black hole. This paper presents a quantitative analysis of trajectory deviations due to mass asymmetry in the two tidal tails. 

In previous studies, either Newtonian or a generalised potential (called `TR' following Tejeda \& Rosswog (2013) \cite{Rosswog_TR}) is used, which is a `low energy limit' (does not mean low velocity) of the true relativistic Schwarzschild dynamics. It is shown that the TR potential accurately mimics orbital dynamics of a test particle as in the exact Schwarzschild background when the orbital distance is large. However, it is acknowledged by the authors that, in the vicinity of the horizon, the tidal interaction will be stronger in the exact Schwarzschild case resulting in a stronger kick velocity. We, in this paper, incorporate exact relativistic acceleration exerted on an object due to a Schwarzschild black hole and perform the simulations for TDEs. 

Trajectory deviations in tidal interactions possess a variety of significance. MacLeod et al. (2013) \cite{MacLeod}, proposed a process of periodic tidal stripping of giant stars as they pass through the pericentre around an SMBH multiple times, thus, feeding to the quiescent luminosity of the SMBH. Our work is important in this context that each pericentre passage may cause tiny deviations in orbits which, after hundreds to thousands of cycles, change the feeding frequency of a black hole significantly. Similarly, this work is important for eccentric tidal disruption studies, e.g., recently performed by Liu et al. (2022) \cite{Liu} about tidal disruption using eccentric Keplerian orbits; study by Cufari et al. \cite{Cufari} on fallback rate after tidal disruption of eccentric orbits. As argued later in this paper, trajectory deviations are prominent, especially in the case of intermediate-mass black holes (IMBHs) ($10^2M_\odot \to 10^5M_\odot$), which can be a significant tool to detect such IMBHs. More importantly, almost three decades long survey and tracking of orbital movements of the S-stars around the Sgr A* SMBH at the center of our galaxy, has opened a new window of observational techniques to decipher nature of black holes, dark matter profile around it as well as stellar properties \cite{ Eckart1996, Ghez1998, Ghez2003, Ghez2008, Schodel2002, Genzel2010, GRAVITY2020, Chan, Becerra-Vergara, Habibi}. Similar observations of stellar orbits around IMBHs will be important for our study where deviation in stellar orbital precession due to tidal effects is required to be considered for a better conclusion from such observations. As argued by \cite{Alexander}, tidal interaction strips the outer layer of a star, initially leaving a redder, more luminous shock-heated remnant core and later a bluer one. Tidally stripped cores, therefore, are expected to be identified observationally and can be tracked to analyse their stellar interior and evolution history.

This paper is organised as follows. In section \ref{sec2}, we give a short review of general relativistic trajectories of a test particle in the Schwarzschild space-time. Next, in section \ref{sec3}, we describe the methodology to simulate the tidal encounters of stellar objects using a smoothed particle hydrodynamics (SPH) code. In section \ref{sec4}, we present the results and provide a quantitative analysis. Finally in section \ref{sec5}, we conclude with a summary and discussion on the significance of our study and future possibilities.   

\section{Exact relativistic acceleration on a point-mass moving in a timelike geodesic in Schwarzschild space-time}
\label{sec2}
Let us consider a point mass moving in a timelike geodesic in Schwarzschild space-time. We want to find the exact relativistic acceleration exerted on an object moving with 4-velocity $ u^\mu $ along its trajectory. We can obtain the constants of motion from the two Killing vectors of the space-time,
 $\xi^\mu_t = \{1,0,0,0\}$ and $\xi^\mu_\varphi = \{0,0,0,1\}$, given as,
\begin{eqnarray}
\epsilon &=& -\xi^t_\mu u^\mu = c^2 \left(1-\frac{2 r_g}{r}\right) \frac{dt}{d\tau} \label{eq_E},\\
l &=& \xi^\varphi_\mu u^\mu = r^2 \frac{d\varphi}{d\tau},
\label{eq_L}
\end{eqnarray}
where, $\epsilon$ and $l$ are the specific orbital energy and the specific OAM respectively. $r_g = G M/c^2 $. Using the above definitions in the 4-momentum conservation of a timelike object $u^\mu u_\mu = -c^2$ we get, 
\begin{equation}
\left(\frac{dr}{d\tau}\right)^2 = \left(\frac{\epsilon^2-c^4}{c^2}\right)-\left[\frac{l^2}{r^2}\left(1-\frac{2 r_g}{r}\right)- \frac{2GM}{r}\right].
\label{eq_rdot}
\end{equation}
Using Equation \ref{eq_E} in Equation \ref{eq_L} and \ref{eq_rdot}, we obtain the velocities with respect to the coordinate time $t$,
\begin{equation}
\dot{\varphi} = \frac{c^2}{\epsilon} \left(1-\frac{2 r_g}{r}\right) \frac{l}{r^2},
\label{eq_dphidt}	
\end{equation} 
\begin{equation}
	\dot{r} = \frac{c^2}{\epsilon} \left(1-\frac{2 r_g}{r}\right) \sqrt{ \left(\frac{\epsilon^2-c^4}{c^2}\right)-\left[\frac{l^2}{r^2}\left(1-\frac{2 r_g}{r}\right)- \frac{2GM}{r}\right]}.
	\label{eq_drdt}
\end{equation}
The `dot' indicates derivative with respect to the coordinate time $t$. It is to be noted that for Schwarzschild metric being a spherically symmetric, static space-time, we can use equatorial plane without any loss of generality, i.e., $\theta = \pi/2$ and $\dot{\theta} = 0 $. Tejeda \& Rosswog define the low energy limit by choosing $\epsilon \sim c^2$ and it leads to the TR potential. However, we are considering the exact relativistic form. An orbit can be determined by correctly choosing the values of the constants $\epsilon$ and $l$. For an elliptical orbit defined by semi-latus rectum $p$ and eccentricity $e$, at the pericentre $r_p = p/(1+e)$ we have $ \dot{r}\vert_{r_p} = 0 $, and also at the apocenter $r_a = p/(1-e)$, $\dot{r}\vert_{r_a} = 0 $. Using Equation \ref{eq_drdt} in these two conditions we obtain,
\begin{eqnarray}
	\epsilon &=& c^2~ \sqrt{\frac{p^2 - 4 p r_g + 4 r_g^2 (1- e^2)}{p^2 - r_g p (e^2 + 3)}},
	\label{Ellipse_E}\\
	l^2 &=& c^2 \left[\frac{r_g p^2}{p -r_g (e^2 +3)}\right].
	\label{Ellipse_L}
\end{eqnarray}
The above equations are also valid for circular ($e=0$), parabolic ($e=1$) and hyperbolic ($e>1$) orbits. It should be noted here that $l$ can attain both $\pm$ sign indicating that the object is moving either in the clockwise or anti-clockwise direction. Now, given the initial position $x_i$ and velocity $\dot{x_i}$ of an object ($i=1,2,3$), we can derive the acceleration $\ddot{x_i} $ on that object from the geodesic equations. Given our choice of the orbit, i.e., $p$ and $e$, we use Equation \ref{Ellipse_E} and \ref{Ellipse_L} to obtain $\epsilon$ and $l$ which are used to obtain the initial velocity components given by Equation \ref{eq_dphidt} and \ref{eq_drdt}. It is convenient to use Cartesian coordinates during the numerical simulation. Therefore, using the initial position and velocity we can obtain the initial acceleration in the Cartesian coordinates as, 
\begin{equation}
\ddot{x_i} = - \frac{G M x_i}{r^3}\left(1-\frac{2 r_g}{r}\right) + \frac{2 r_g \dot{x_i}\dot{r}}{r(r-2 r_g)} + \frac{r_g x_i \dot{r}^2}{(r- 2 r_g)r^2} - \frac{2 r_g x_i \dot{\varphi}^2}{r}.
\label{Sch_acceleration}
\end{equation}  
Here, the conversion of $\dot{r}$ and $\dot{\varphi}$ from polar to the Cartesian coordinates are 
\begin{eqnarray}
r &=& \sqrt{x^2 + y^2 + z^2}, \\
r\dot{r} &=& x \dot{x} + y \dot{y} + z\dot{z},\\
r^4 \dot{\varphi}^2 &=& (x \dot{y} - y \dot{x})^2 +  (x \dot{z} - z \dot{x})^2 +  (z \dot{y} - y \dot{z})^2.   
\end{eqnarray}
We are now capable of integrating Equation \ref{Sch_acceleration} using the initial position and velocity of an object and find its exact general relativistic trajectory in the Schwarzschild space-time. It is important to note here that for an extended object having a finite volume, each infinitesimal volume element of the object is exerted upon by the external acceleration given by Equation \ref{Sch_acceleration}. Consequently, tidal force automatically appears when forces on different parts of the object are compared on the macroscopic scale. 

\section{Methodology}
\label{sec3}
\subsection{Details of the code}
To simulate tidal disruption, we have developed a 3D fast, parallel code based on smoothed particle hydrodynamics (SPH). The code is written in C following closely the algorithm described in PHANTOM \cite{PHANTOM} (also see \cite{SEREN}, \cite{Rosswog_tree}). In SPH, a fluid element is considered as a particle. Thus a fluid star is formed using a collection of particles. Density, pressure and other hydrodynamic properties of a particle are calculated from its neighbour particles. A binary tree is implemented for efficient neighbour searching. In a binary tree, a group of particles (node) are bisected at their center of mass (COM) position along the longest dimension to obtain two children nodes. This procedure is repeated until the whole system is covered by lowest level cells having at most $N_{ll} = 10$ particles. For a system containing $N$ particles, a binary tree brings down the expense of neighbour search from $\mathcal{O}(N^2)$ to $\mathcal{O}(N\log N)$. The discrete nature of particles are smoothed out using spherically symmetric M6 quintic spline kernel while calculating hydrodynamic properties and forces.

The self-gravity of the system is calculated in two steps: near field and far field gravity. Near field gravity on a particle due to its neighbours, is calculated using a gravitational kernel having the same kernel radius as M6. On the other hand, when a node is far away from the particle of interest, its gravitational force is calculated using its monopole and quadrupole moments. A multipole acceptance criteria (MAC) is used to determine whether a node B contributes to the far field of node A. A tree opening angle $\theta = 0.5 $ is used throughout this work. Node B falls under the far field of node A if the MAC condition holds: 
\begin{equation}
\theta \ge \frac{L^B_{max}}{R_{AB}},
\end{equation}       
where, $L^B_{max} = max \{L_x,L_y,L_z\}$ is the maximum of the side lengths of the node B along three spatial dimensions, and $R_{AB}$ is the distance between the COM of node A and B. The relative error in acceleration for using $\theta=0.5$ is less than $0.1\%$, whereas for $\theta =1$ it is $\sim 1\%$. The external gravity, exerted on each particle, due to the black hole is calculated using the relativistic acceleration in Schwarzschild space-time given by Equation \ref{Sch_acceleration}. 

For accurate shock capturing, we have incorporated artificial viscosity (AV) in the code. Each particle is assigned its own value of artificial viscosity using the version of Price \cite{Price_AV} and Rosswog et al. \cite{Rosswog_AV} improved from the Morris Monaghan method \cite{MM1997}. Individual and temporally adaptive viscosity enables us to increase its strength quickly enough (up to a maximum limit) to capture convergent flows or a pre-shock region accurately. In presence of divergent flows or at post-shock region, its strength decays down to a minimum value set by the user. In this work to simulate TDEs, however, we have used constant values of the AV parameters.

The temporal evolution of the particle system is done using the Kick-Drift-Kick (KDK) leapfrog integration scheme. KDK method is a time-reversible, and symplectic integration scheme \cite{Gadget, Hairer}, which preserves the Hamiltonian nature of the SPH formalism. Total energy, linear and angular momentum of the system remain conserved. In KDK approach, at first, acceleration is calculated (Kick) at time step $t_n$, which is used to evolve the velocity at the half time step $t_{n+1/2}$. Next, position is evolved from $t_n$ upto the full time step $t_{n+1}$ (Drift). Finally, new acceleration is calculated (Kick) at the updated position, which is used to update the velocity at $t_{n+3/2}$. In case of velocity dependent acceleration, as in Equation \ref{Sch_acceleration}, we make an extrapolation of the velocity from $t_{n+1/2}$ to $t_{n+1}$ to be used to calculate acceleration at $t_{n+1}$. In the latter case, the final `Kick' is repeated until the velocity used for calculating acceleration and the corrected velocity at $t_{n+1}$ converge to a prescribed accuracy having relative error less than $10^{-2}$. 

The time step for evolution is chosen in the following two ways. In global timestepping scheme, all particles are evolved using the smallest time step of all the particles, $dt_{min}$. In order to make the time evolution efficient, we have implemented individual time stepping \cite{indiv_timestep}. Each particle evolves according to its own time step which is readjusted to $ dt' = 2^n dt_{min} $ from its actual time step $dt$, where $n$ is the maximum integer such that $dt' \leq dt $. Particles having the same $n$ correspond to the same time bin. Time step of the $i+1$th bin is half of the $i$th bin's time step. Because of the bin construction, time evolution of the whole system does not slow down when very few particles achieve extremely small time steps. However, in few extreme situations (e.g., Sedov blast, blast wave) when neighbour particles are separated by multiple bins, the slowly evolving particles cannot react quick enough to their rapidly evolving neighbours. To overcome this issue, a `wake up' scheme is implemented such that neighbours are always kept in adjacent bins. Energy conservation is maintained within $0.2\%$ in case of global timestepping as compared to $3\%$ in individual timestepping. In this paper, global timestepping is used to prefer accuracy over efficiency.

We have incorporated hybrid parallelisation using OpenMP and MPI. Our code is verified with standard tests, e.g., Sod shock tube, Kelvin-Helmholtz instability, colliding flow test, Sedov blast, Einfeldt test, Evrard collapse, polytropic star formation, radial oscillation etc. The simulation results agree with the analytical outcome and existing literature standards. All the simulation results presented in this paper are obtained with $N=5\times10^5$ particles unless otherwise stated. Column density plots are produced from the simulation data using the visualisation tool SPLASH \cite{Price_SPLASH}.

\subsection{Simulation of tidal disruption}
\subsubsection{Initial condition}
We want to evolve a polytropic fluid star which initially is kept at a distance far away from the tidal influence of the black hole. When a star with mass $M_\star$ and radius $R_\star$ approaches near a black hole of mass $M$, it is tidally disrupted if the tidal force at the stellar surface overcomes its self gravity. We define a tidal radius $r_t$ from the black hole as:
\begin{equation}
	r_t = \left(\frac{M}{M_\star}\right)^{1/3} R_\star,
	\label{tidal_formula}
\end{equation}
which crudely means that the star is tidally disrupted if it enters within $r_t$.
In this work, we initially place a spherically symmetric polytropic fluid star with $M_\star = 1 M_\odot$ and $R_\star = 1 R_\odot$ and polytropic index $n=1.5$ at a distance of $3r_t$ where the tidal forces are practically negligible. The initial density distribution should be equivalent to the solution of the Lane Emden equation for the chosen polytrope. Required particle distribution is obtained using stretch mapping technique \cite{Herant}. Starting from a sphere of uniform HCP particle distribution of radius $r_{max}$, stretch mapping creates the desired Lane Emden density profile. A particle having radial distance $r_0$ from the center of the sphere is `stretched' to a new radial position $r$, such that
\begin{equation}
	\frac{M_{_{LE}}(r)}{M_{_{LE}}(r_{max})}=\frac{r_0^3}{r_{max}^3},
\end{equation}
where $M_{_{LE}}(r)$ is the desired mass at radius $r$ of the Lane Emden density profile. The stretch-mapped particle distribution is then evolved such that the particles can settle to achieve almost perfect Lane Emden density profile. It is now ready for further evolution in presence of the external gravity of a black hole. We are still required to specify the initial velocity of the star such that it follows the desired trajectory. We choose parabolic orbits with eccentricity $e=1$, while $p$ is determined by our choice of pericentre $r_p$. We find the initial velocity from Equation \ref{eq_dphidt} and \ref{eq_drdt} and convert them into Cartesian coordinates. Each particle inside the star is given the same initial velocity such that the star moves along the parabolic trajectory.

\subsubsection{Estimation of the bound core}

As the fluid star moves near $r_t$, the tidal force gets stronger and deforms the stellar body. Here, we define a parameter $\beta = r_t/r_p$ to determine the amount of tidal disruption. If $ \beta \lesssim 1$, the star is partially disrupted where the outer layer gets stripped away and the remaining core recoils back due to its self-gravity. On the other hand, for $\beta \gtrsim 1$ the star enters within the tidal radius and gets completely disrupted. In this paper,  we intend to study the dynamics of the remnant core. Therefore, it is required to calculate the specific orbital energy and OAM of the core. The process is described in the following.
    
To identify those particles that are bound to the core, we follow an energy based iterative method introduced by Guillochon \& Ramirez-Ruiz \cite{Guillochon2013}. To begin with, the particle having the peak density is identified and its velocity $\textbf{v}_{\text{peak}}$ is used to calculate the specific binding energy of another particle $i$ as
\begin{equation}
	\varepsilon_i = \frac{1}{2} \left(\textbf{v}_i - \textbf{v}_{\text{peak}}\right)^2 - \phi_i,
	\label{binding_energy}
\end{equation}
where, $\textbf{v}_i$ denotes the velocity of the $i$th particle. $\phi_i$ is the self gravitational potential. Particles having $\varepsilon_i < 0$ are bound to the core. Therefore, the center of mass velocity of the bound core is obtained as
\begin{equation}
	\textbf{v}_{\text{core}} = \frac{\Sigma_{\varepsilon_i<0}~ \textbf{v}_i m_i }{\Sigma_{\varepsilon_i<0}~ m_i}.
	\label{eq_vcm}
\end{equation}
$\textbf{v}_{\text{core}}$ is replaced in place of $\textbf{v}_{\text{peak}}$ in Equation \ref{binding_energy}, and the process is repeated until $\textbf{v}_{\text{core}}$ converges. Finally, adding mass of all the bound particles, we get the bound core mass $m_{\text{core}}$. Similar to Equation \ref{eq_vcm}, we find the center of mass position $\textbf{r}_{\text{core}} $. Note that, using the position and velocity of a particle we can get its $\epsilon$ and $l$ from Equation \ref{eq_dphidt} and \ref{eq_drdt}. Similarly, using $\textbf{v}_{\text{core}}$ and $\textbf{r}_{\text{core}}$, we obtain the specific orbital energy $\epsilon_{\text{core}}$ and angular momentum $l_{\text{core}}$ of the core, a change of which leads to the deviation of its trajectory.

\section{Simulation results and analysis}
\label{sec4}

In this work, we send a solar-type star ($1 M_\odot$, $1 R_\odot$, $n =1.5$) around a black hole having $M =100 M_\odot$ in different parabolic trajectories starting from $\beta = 0.6$ to $ 0.9 $ in intervals of $0.05$. The range of $\beta$ is chosen such that partial tidal disruption is ensured and a remnant bound core is formed. Moreover, the black hole mass is chosen in order to generate asymmetry in the tidal force fields and mass difference between the two tidal tails. Asymmetry appears when the stellar radius $R_\star$ and pericentre $r_p$ are comparable to $r_t$. As evident from Equation \ref{tidal_formula}, $M$ needs to be $\mathcal{O}(10^3 M_\odot)$ or less for  $r_t \lesssim 10 R_\star$. Therefore, for the aforementioned range of $\beta$, we have $11 R_\star \lesssim r_p \lesssim 17 R_\star $. It is important to note here that tidal asymmetry increases with increasing $\beta$. By choosing $M=100M_\odot$ for this work and consequently, $r_t = 4.64 R_\star$, and $5 R_\star \lesssim r_p \lesssim 8 R_\star $, we have ensured that the influence of asymmetry on trajectory deviation is prominent.
\begin{figure}[h!]
	\centering 
	\includegraphics[scale=0.25]{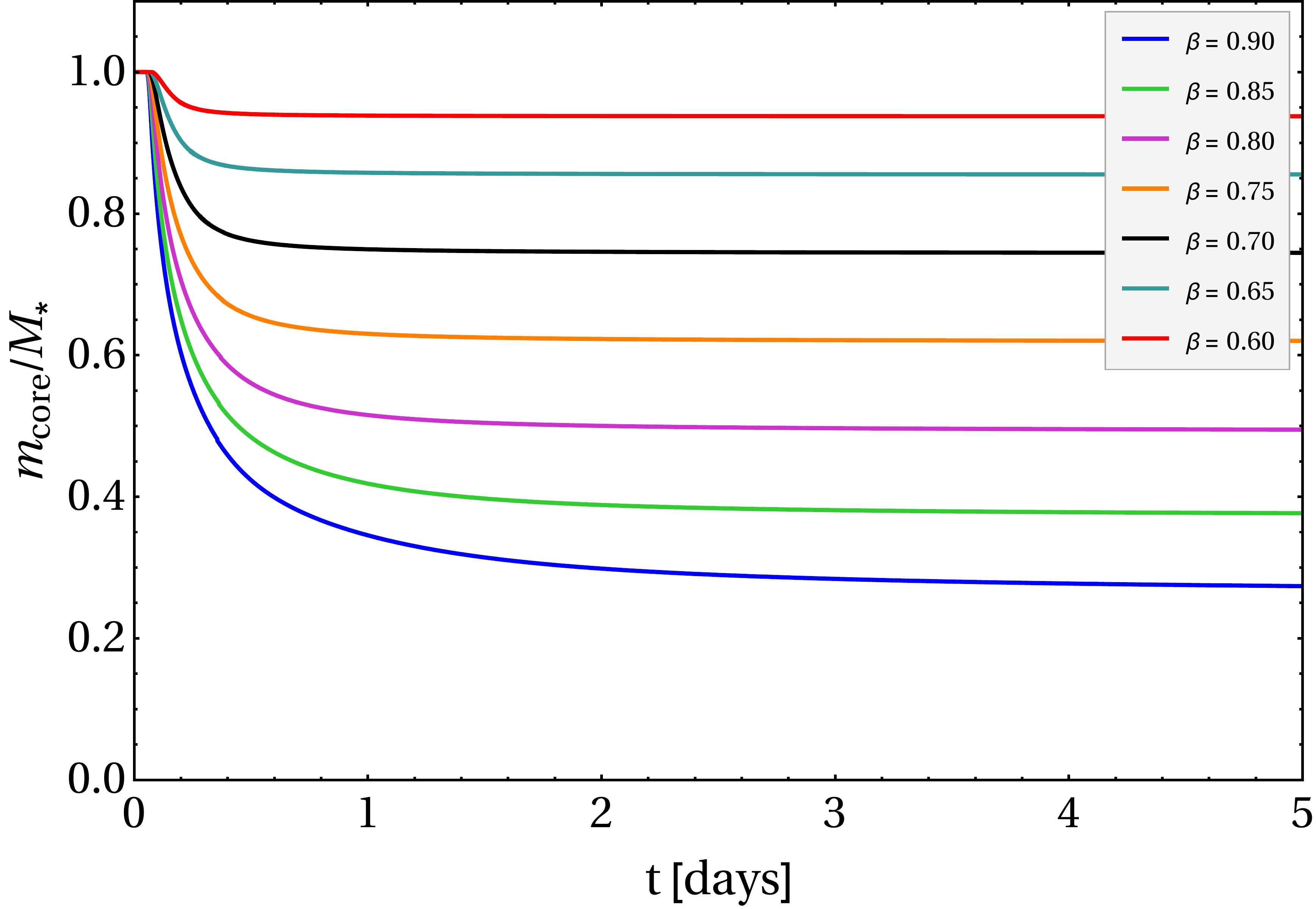}
	\includegraphics[scale=0.25]{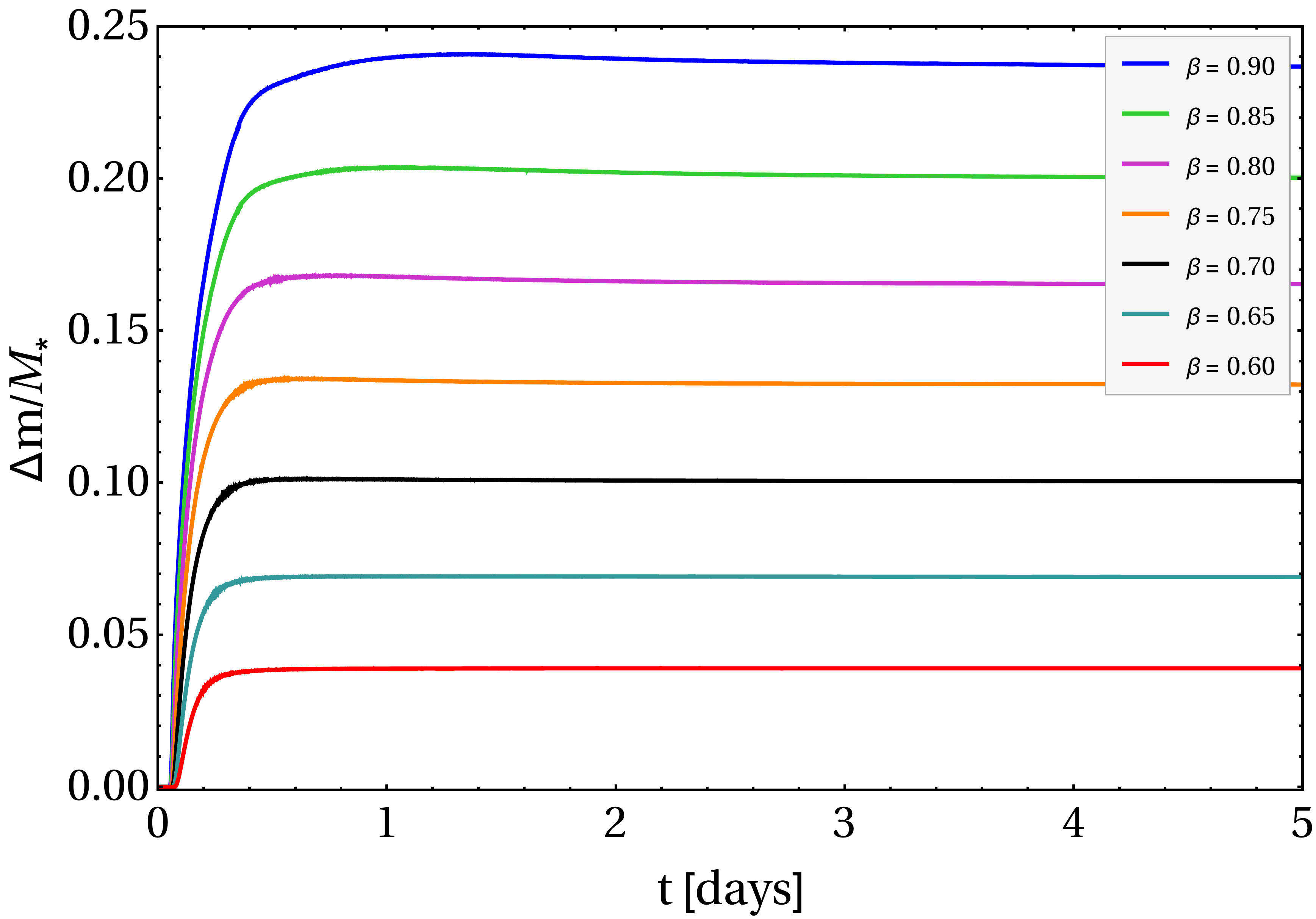}
	\caption{\textbf{Left panel} Mass fraction of the remnant self-bound core $m_{\text{core}}$ with respect to its initial mass $M_{\star}= 1M_\odot$ is plotted with time for various values of $\beta = 0.6$ to $0.9$ in intervals of $0.05$. Higher $\beta$ causes stronger tidal interaction which leads to higher mass loss. \textbf{Right panel:} Time evolution of the mass difference between the two tidal tails $\Delta m$ is presented. With higher $\beta$, asymmetry increases and so does the mass difference.}
	\label{fig.mcore}
\end{figure}
\begin{figure}[h!]
	\centering
	\includegraphics[scale=0.245]{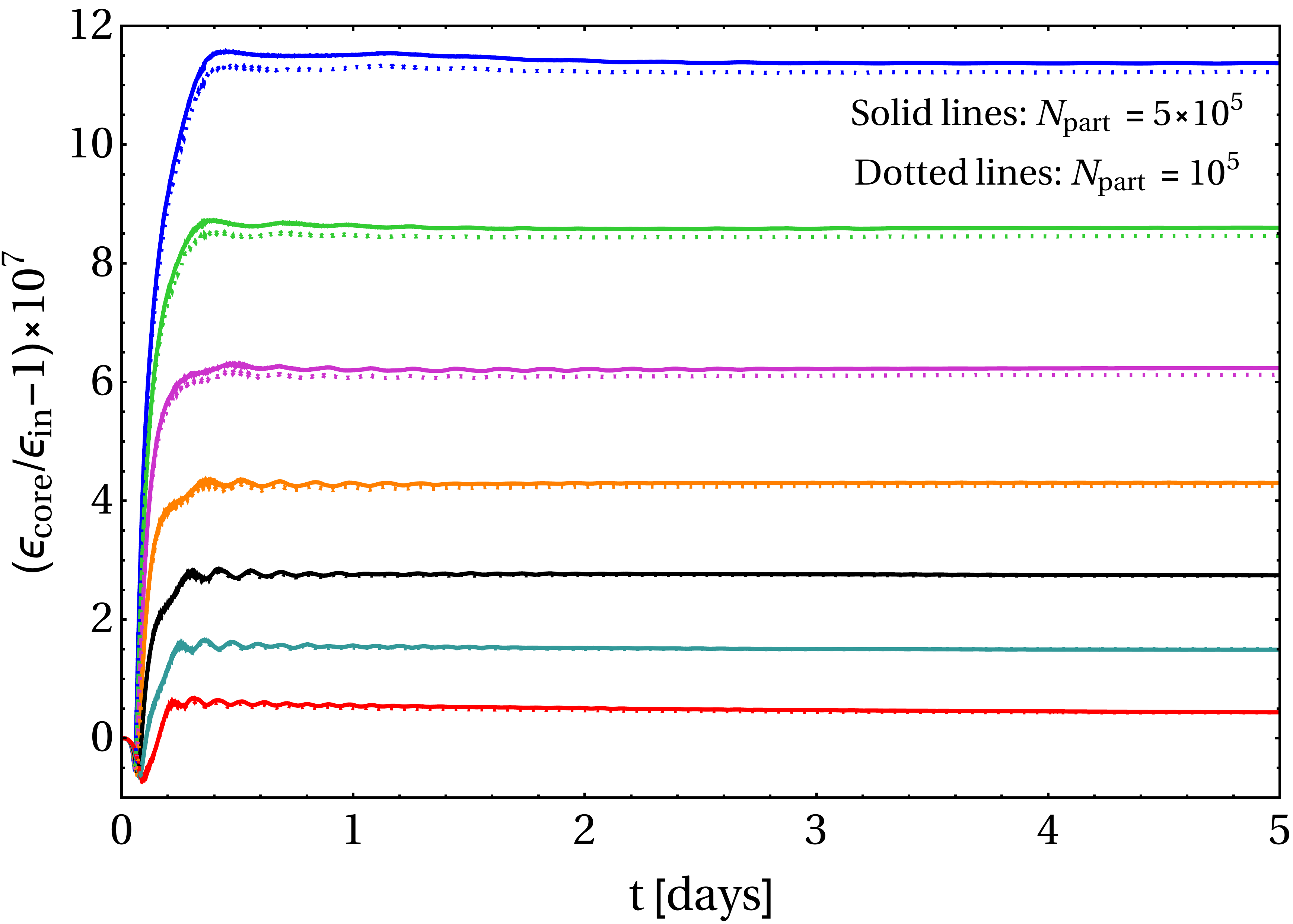}
	\includegraphics[scale=0.245]{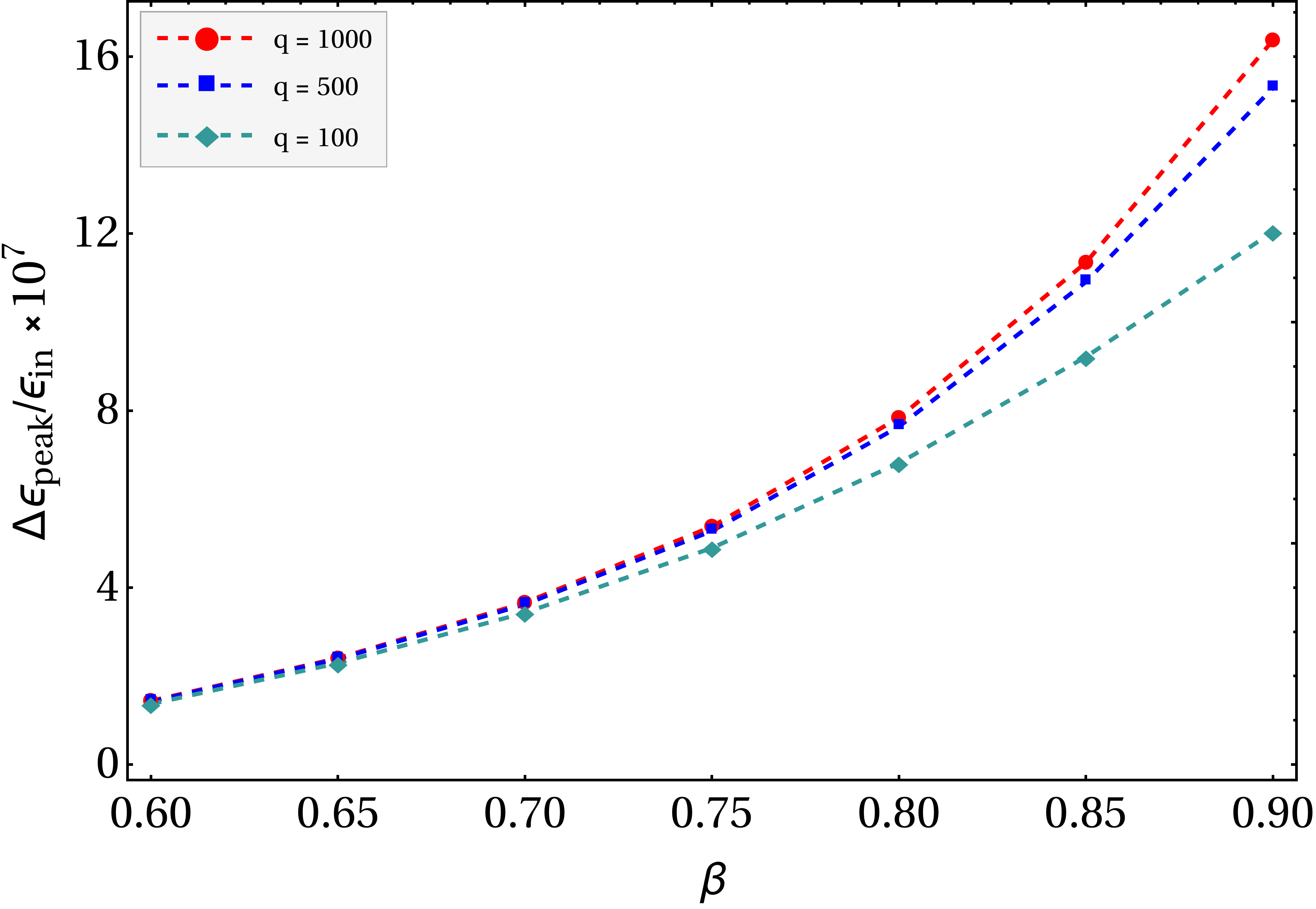}
	\includegraphics[scale=0.25]{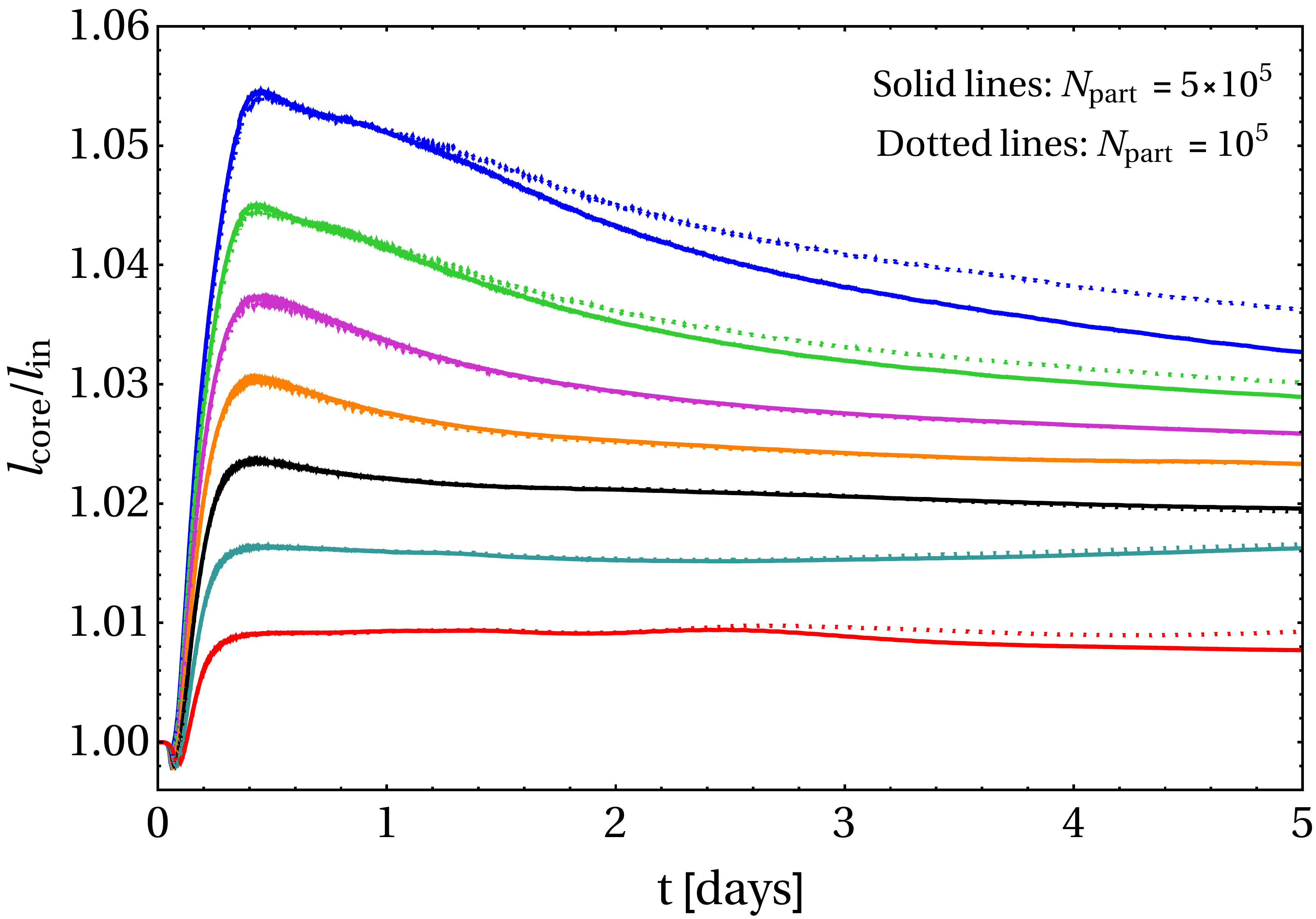}
	\includegraphics[scale=0.25]{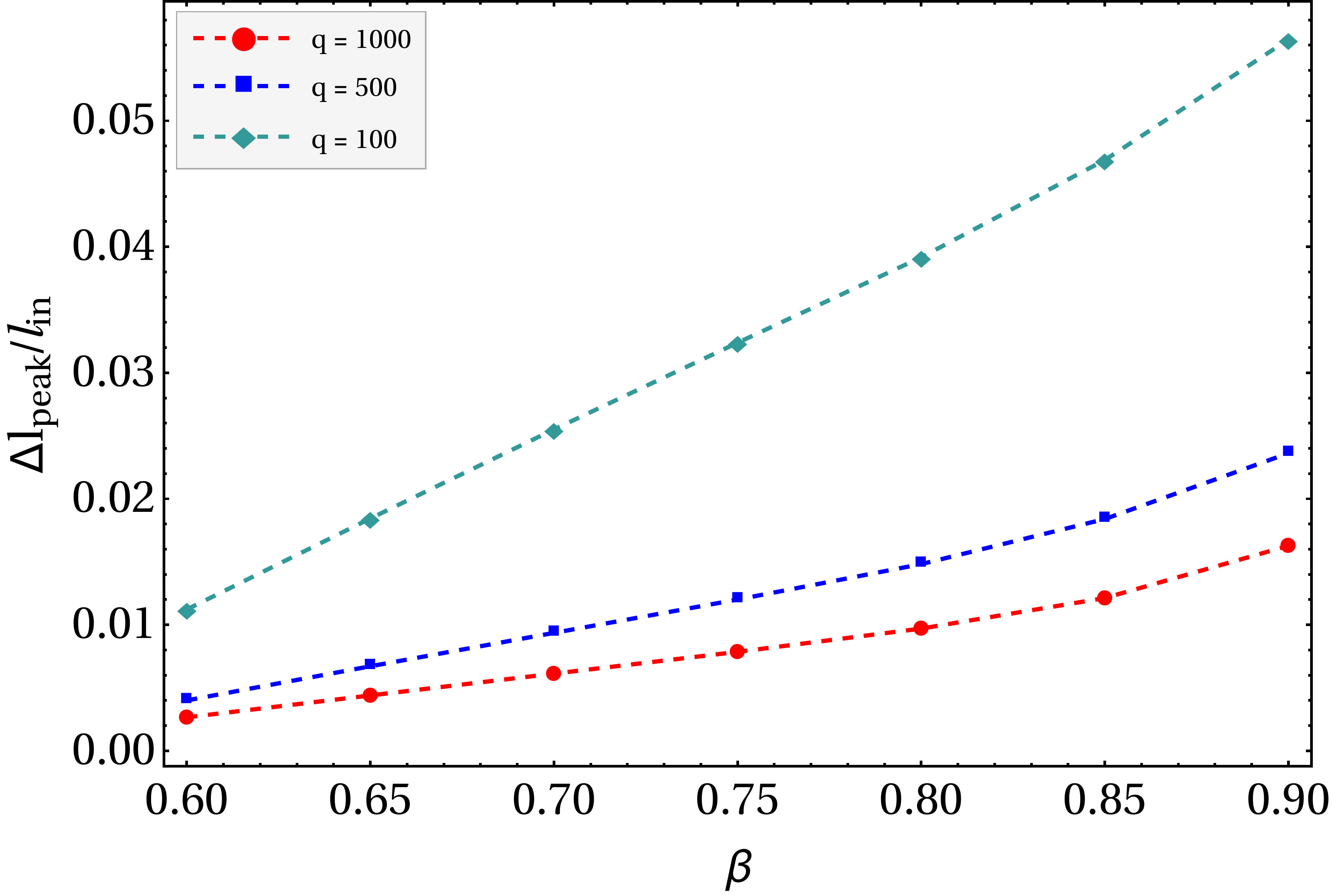}	
	\caption{\textbf{Top left panel:} Relative deviation in orbital energy of the self-bound core $\epsilon_{\text{core}}$ from its initial value $\epsilon_{\text{in}}$ is plotted with time for various values of $\beta = 0.6$ to $0.9$ in interval of $0.05$. \textbf{Top right panel:} Peak relative deviation in the orbital energy of the core as a function of $\beta$ for 3 different mass ratios $q = 100,500, 1000$. \textbf{Bottom left panel:} Specific orbital angular momentum of the core $l_{\text{core}}$ as normalised to its intitial value $l_{\text{in}}$ is plotted with time for $\beta = 0.6$ to $0.9$ in intervals of $0.05$. \textbf{Bottom right panel:} Peak deviation in $l_{\text{core}}/l_{\text{in}}$ increases with $\beta$, as shown for three $q$ values.}
	\label{fig.eob_time}
\end{figure}

\subsection{Bound core mass-fraction}
After the solar-type star is partially disrupted, stellar debris leaves the star's gravitational influence via the L1 (towards the black hole) and L2 (away from the black hole) Lagrange points. As the star approaches in a parabolic orbit, the center of mass of the star has specific orbital energy $c^2$ (also evident from Equation \ref{Ellipse_E}). The material ejected from L1, loses orbital energy ($< c^2$) and forms the bound tidal tail, whereas the other tail formed via L2, gains orbital energy ($>c^2$) to become unbound. The bound tail eventually falls back into the black hole and gets accreted. On the other hand, the unbound tail never returns. See \cite{Cufari} for a discussion on the boundedness of tidal tails for elliptical and hyperbolic orbits. In case of partial disruption, the outer layer of the star is disrupted while the remnant core gradually recoils back to its spherical shape owing to its self-gravity as it moves away from the black hole after the tidal encounter. The mass of the bound core saturates at a constant value; similarly, the bound and unbound tails gradually get disconnected from the core. Figure \ref{fig.mcore} left panel shows that the core mass fraction decreases with time. Due to the asymmetry in the tidal force fields near the star, more mass is ejected from L1 towards the black hole than L2, leading to a mass difference between the tidal tails as clearly depicted in the right panel of Figure \ref{fig.mcore}. As the system evolves, the mass difference between the two tidal tails increases. As expected, for higher $\beta$, higher asymmetry leads to a higher mass difference in the tidal tails. 

\subsection{Variation of specific orbital energy and specific orbital angular momentum}
Due to the conservation of linear momentum, additional momentum carried away by the bound tail imparts a `kick' on the remnant self-bound core, increases its velocity, and deviates the core from its initial trajectory. In this process, the specific orbital energy $\epsilon_{\text{core}}$ and angular momentum $l_{\text{core}}$ of the core increases due to the tidal interaction. The bound core turns into a hypervelocity star (see \cite{Manukian, Rosswog_tidalkick}). In Figure \ref{fig.eob_time} top left panel, deviation in specific orbital energy of the core $\Delta\epsilon_{\text{core}} = \epsilon_{\text{core}} - \epsilon_{\text{in}}$, normalised to its initial value $\epsilon_{\text{in}}= c^2$ is plotted with time for multiple values of $\beta$. As the star approaches the pericentre, pure asymmetric deformation of the stellar body decreases the specific orbital energy of the whole star resulting in an initial dip in the plot. As the outer layer starts to disintegrate and forms the tidal tails, mass difference between the tails further leads to the kick of the bound core. Consequently, $\epsilon_{\text{core}}$ rapidly jumps to its maximum. Eventually, as the mass of the bound core saturates, so does its $\epsilon_{\text{core}}$. The jump in the specific orbital energy of the core $\Delta\epsilon_{\text{peak}} = \text{max}(\epsilon_{\text{core}}) - \text{min}(\epsilon_{\text{core}})$ is shown in the top right panel of Figure \ref{fig.eob_time}. It is evident that with higher $\beta$, increased asymmetry in the tidal tails causes a higher jump.

Initially, during the pure asymmetric tidal deformation phase, the stellar material experiences strong tidal forces. Therefore, consider a fluid element which drives away from the COM trajectory towards the black hole, such that the angular velocity ($\dot{\varphi}$) of the fluid element decreases from that of the COM. From Equation \ref{eq_L} we see that the specific OAM of the fluid element also decreases. As more stellar material is deformed towards the black hole due to the asymmetric tidal force fields, the specific OAM of the deformed star decreases. This is evident from the initial dip in Figure \ref{fig.eob_time} (bottom left panel). As the star passes through the pericentre, and two tidal tails form, the bound tail moves towards the black hole and loses OAM while the unbound tail gains it. Due to the mass difference in the two tails, the core gains specific OAM and reaches its peak value. From the figure, $l_{\text{core}}$ is found to be gradually decreasing from its peak for higher values of $\beta$. It suggests that the tidal tails are not yet completely separated from the remnant core, and OAM transfer continues such that $l_{\text{core}}$ decays before eventually achieving a constant value. At this point, the bound tail forms an accretion disk, and the core gains density. It leads to extremely small timesteps making the simulation slow. Therefore, we do not continue beyond five days.

\begin{figure}[h!]
	\centering
	\includegraphics[scale=0.27]{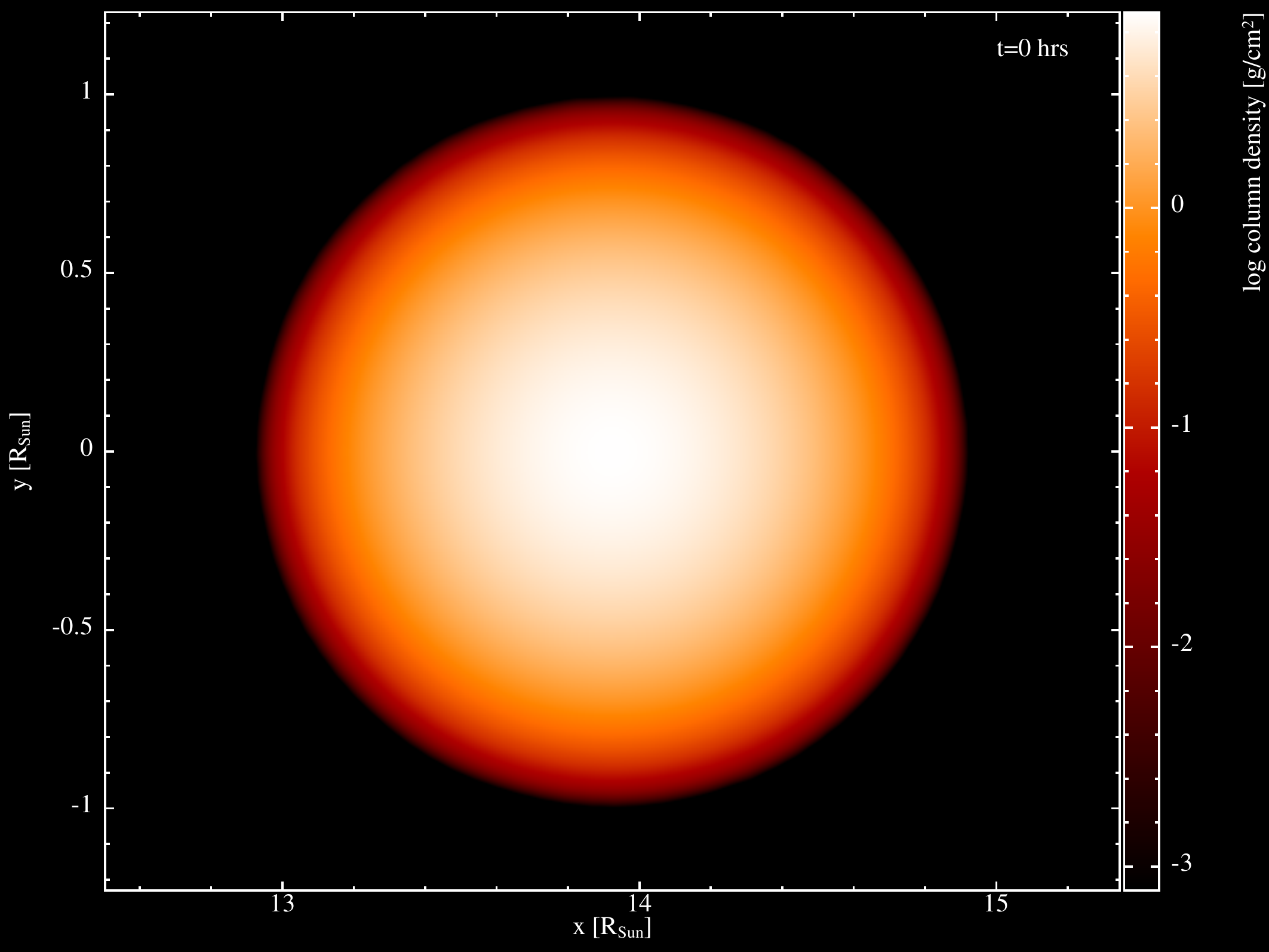}
	\includegraphics[scale=0.27]{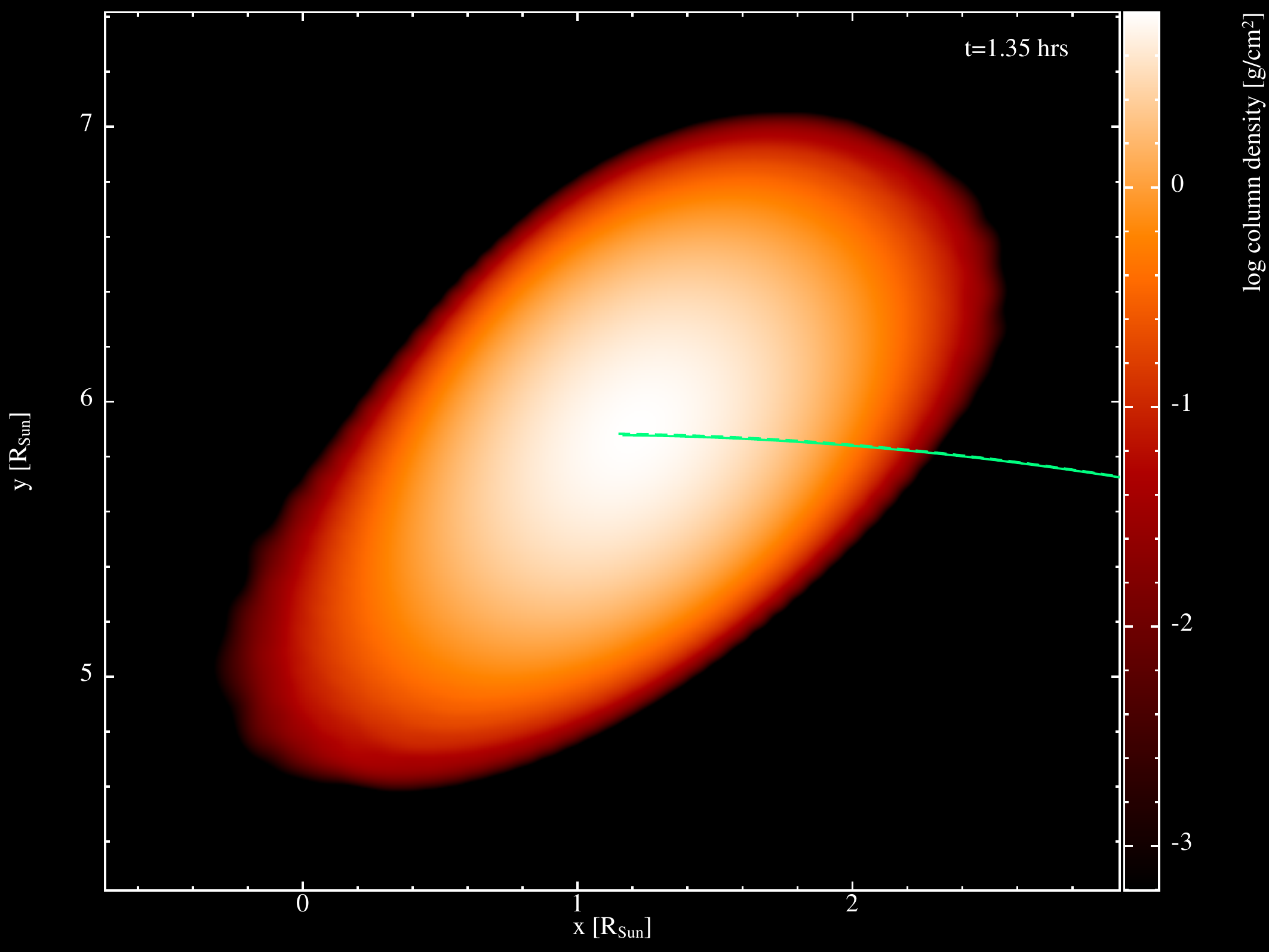}
	\includegraphics[scale=0.27]{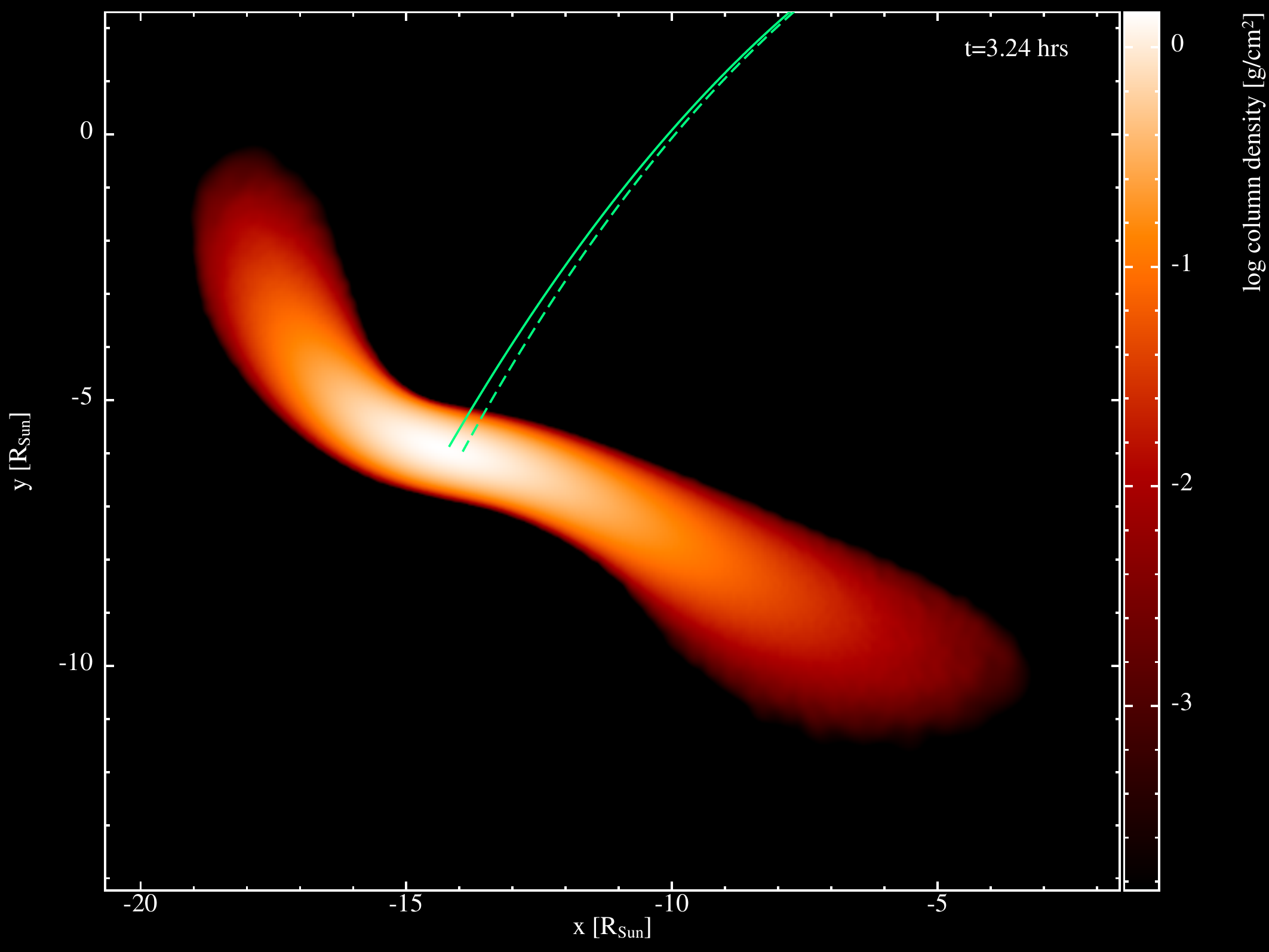}
	\includegraphics[scale=0.27]{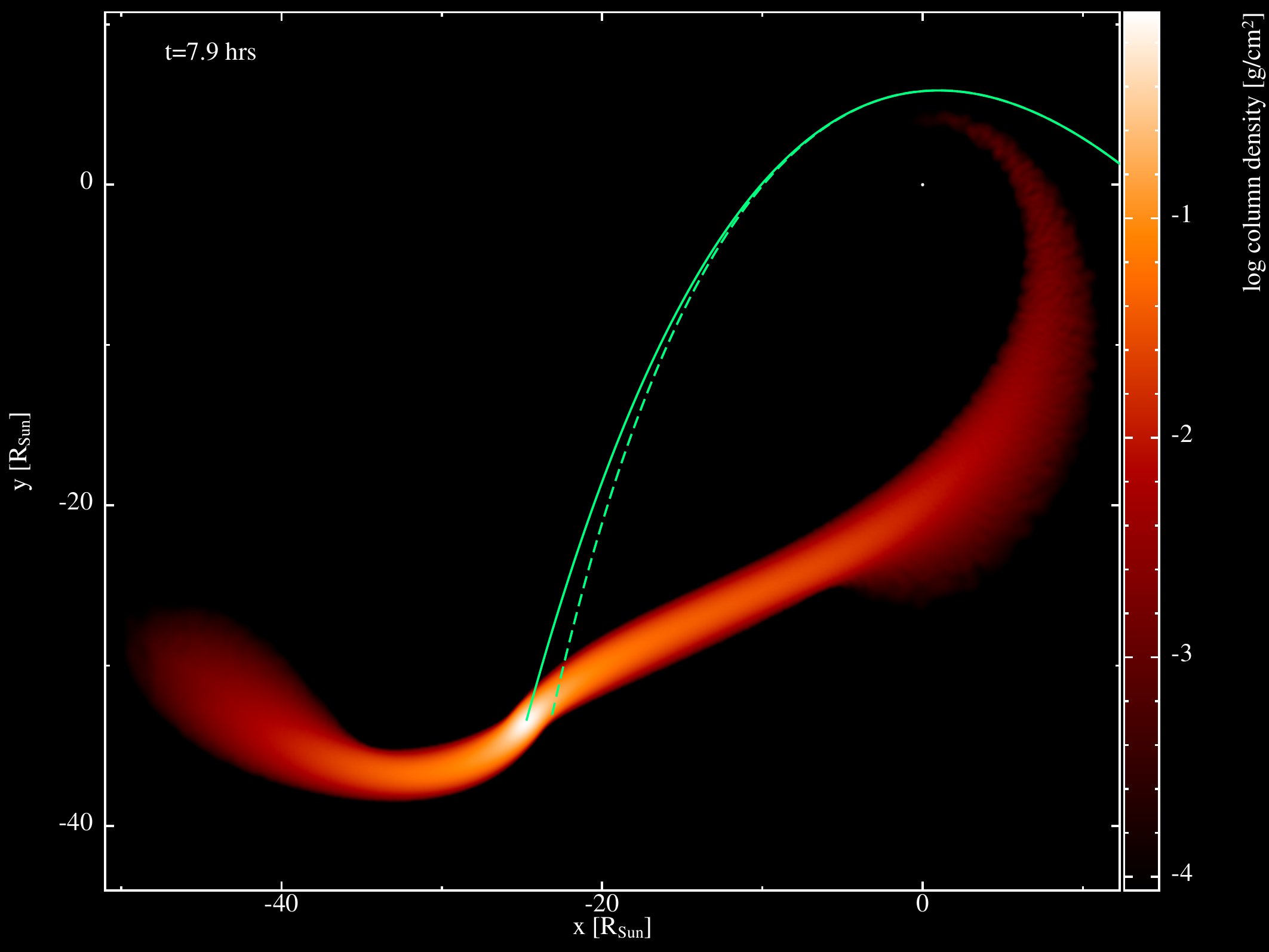}
	\includegraphics[scale=0.27]{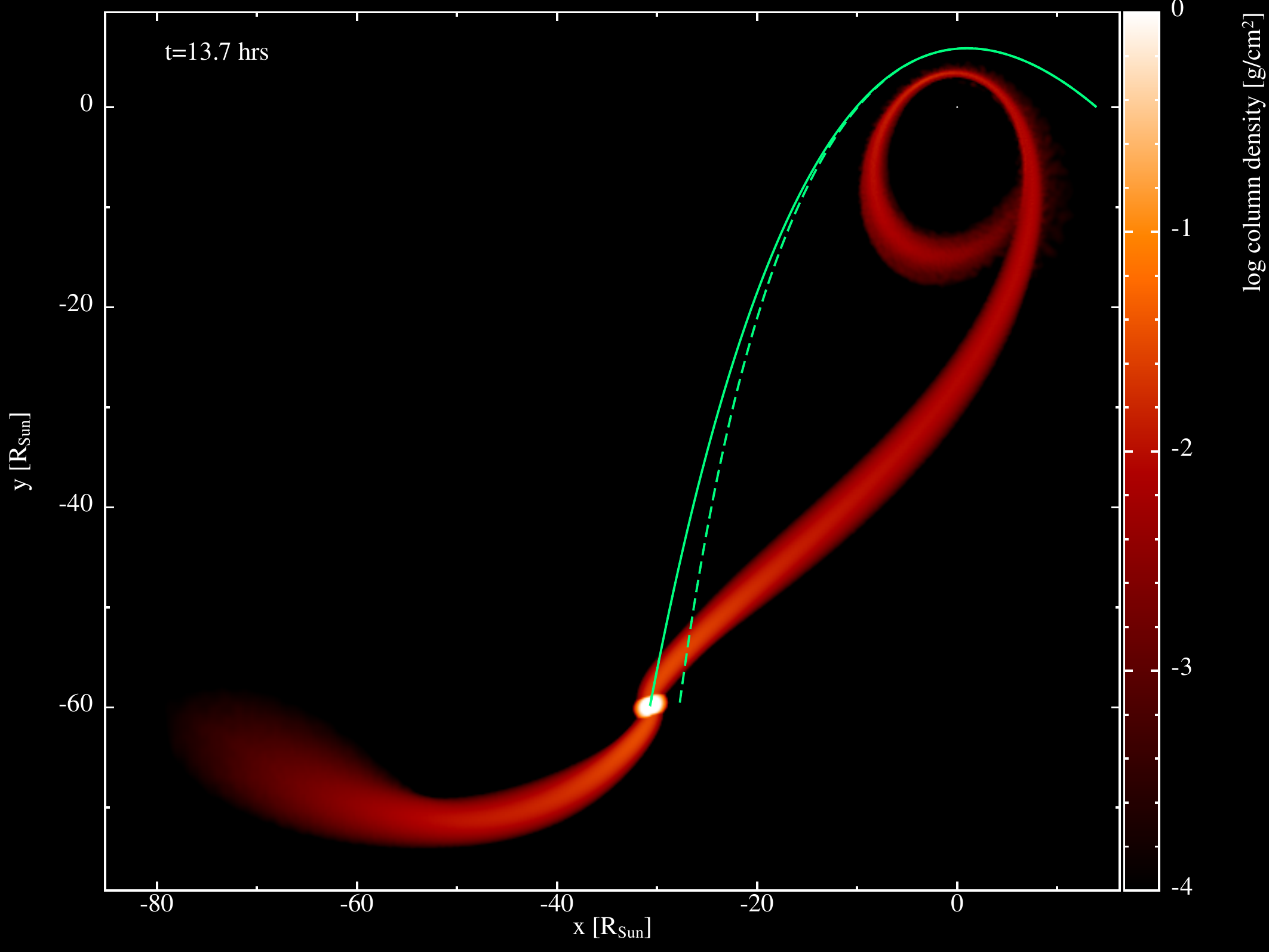}
	\includegraphics[scale=0.27]{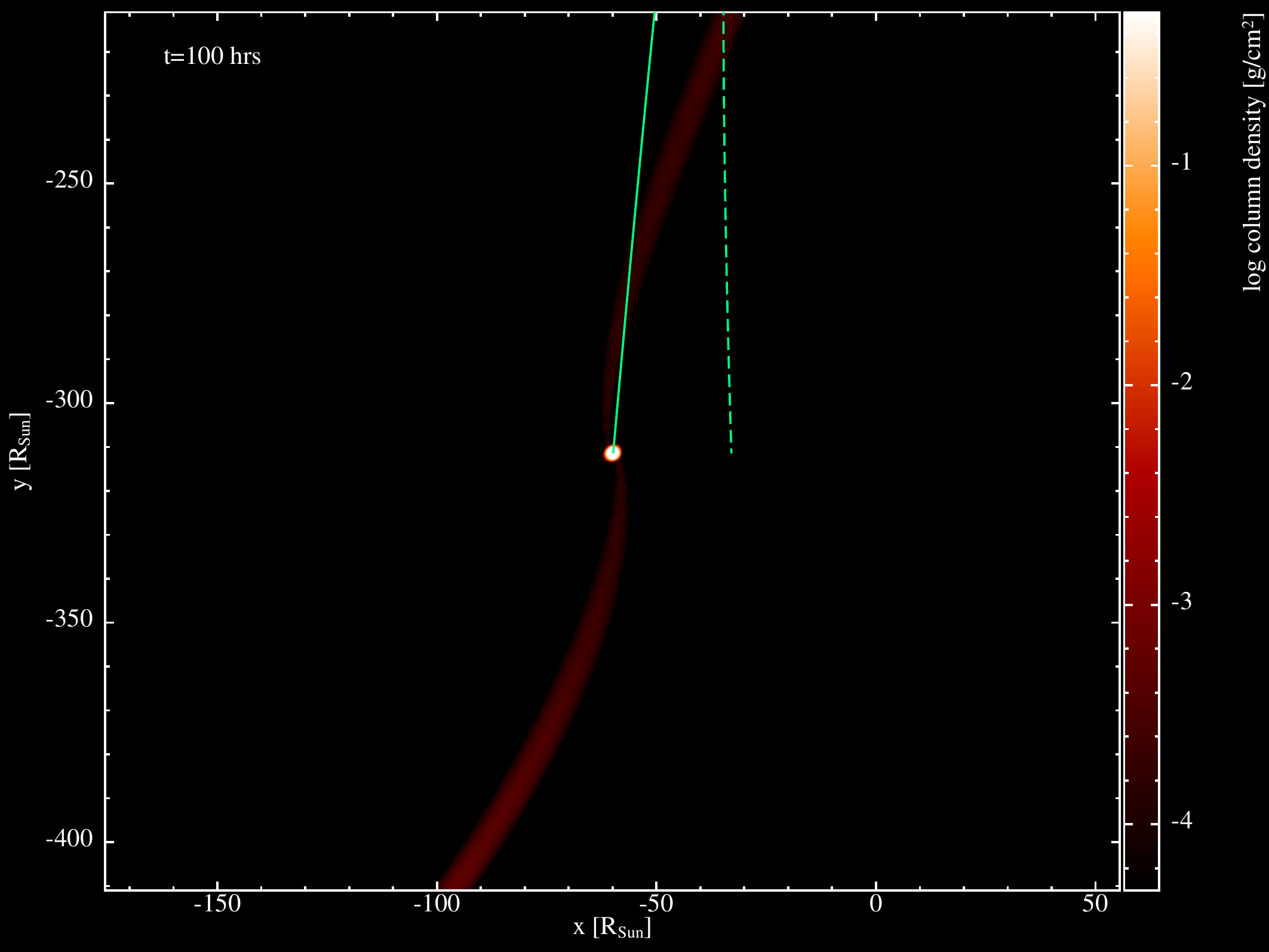}

	\caption{Column density plots of the star having mass $M_\star = 1 M_\odot$ , radius $R_\star = 1 R_\odot$ and polytropic index $n=1.5$ during its tidal interaction with $\beta = 0.8$ with a Schwarzschild black hole of $M =100M_\odot$ placed at $(0,0,0)$ (marked as white dot). Solid green line indicates the actual trajectory of the remnant core up to its current position. Dashed green line demonstrates the initial parabolic trajectory along which the star approaches the black hole. \textbf{Top left panel:} Initial spherical shape of the star at a distance $3r_t$. \textbf{Top middle panel:} Pure asymmetric deformation at pericentre. \textbf{Top right panel:} Tidal tail formation starts. \textbf{Bottom left panel:} Two tidal tails with different masses. \textbf{Bottom middle panel:} Remnant self-bound core is visible. \textbf{Bottom right panel:} After 100hrs, the remnant core loses almost $60\%$ of its initial mass, and is deviated from its initial parabolic trajectory. }
	\label{fig.density}
\end{figure}

In the bottom right panel of Figure \ref{fig.eob_time}, peak deviation in $l_{\text{core}}$, denoted as $\Delta l_{\text{peak}}$, is plotted relative to its initial value $l_{\text{in}}$ with $\beta$ for three mass ratios $q= M/M_\star = 100, 500$ and $1000$. As expected, with higher $\beta$ for a fixed $q$, tidal asymmetry increases and consequently, peak deviation in $l_{\text{core}}$ increases. On the other hand, tidal asymmetry decreases for higher $q$, as already discussed at the beginning of this section. Therefore, $\Delta l_{\text{peak}}/l_{\text{in}}$ is found be lower. Surprisingly, from the top right panel, we see that relative deviation in specific orbital energy is higher for higher $q$. The reason is that the initial specific orbital energy is $c^2$ for all parabolic orbits irrespective of black hole mass $M$ or mass ratio $q$ and $\beta$. Therefore, following the absolute increment in specific orbital energy, relative deviation also increases. On the other hand, initial specific OAM of the star moving in a parabolic orbit with the same $\beta$, increases as $\propto M^{2/3}$. It causes the relative deviation in specific OAM to decrease with $M$ or $q$. 

\begin{figure}[h!]
	\centering	
	\includegraphics[scale=0.25]{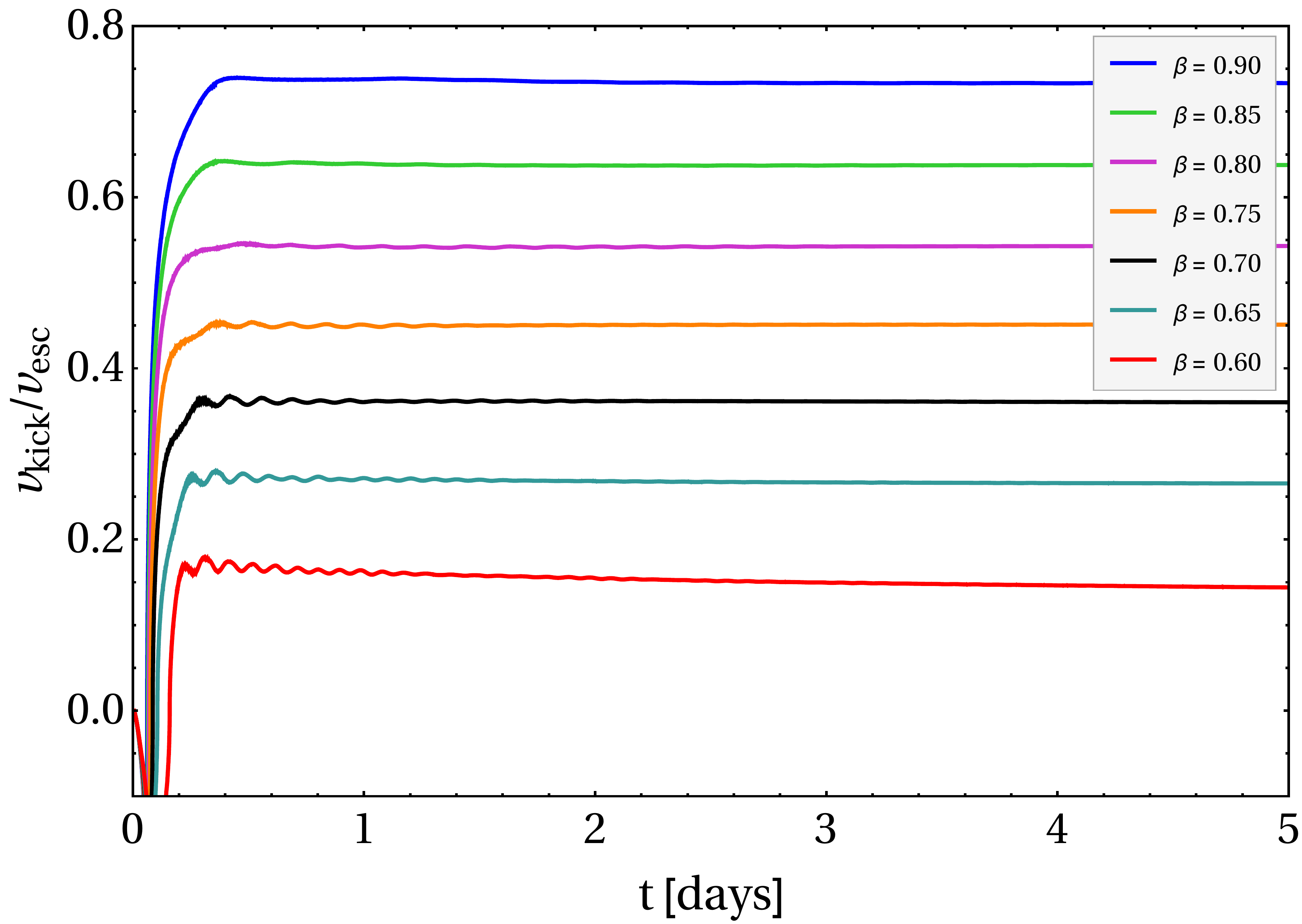}
	\includegraphics[scale=0.25]{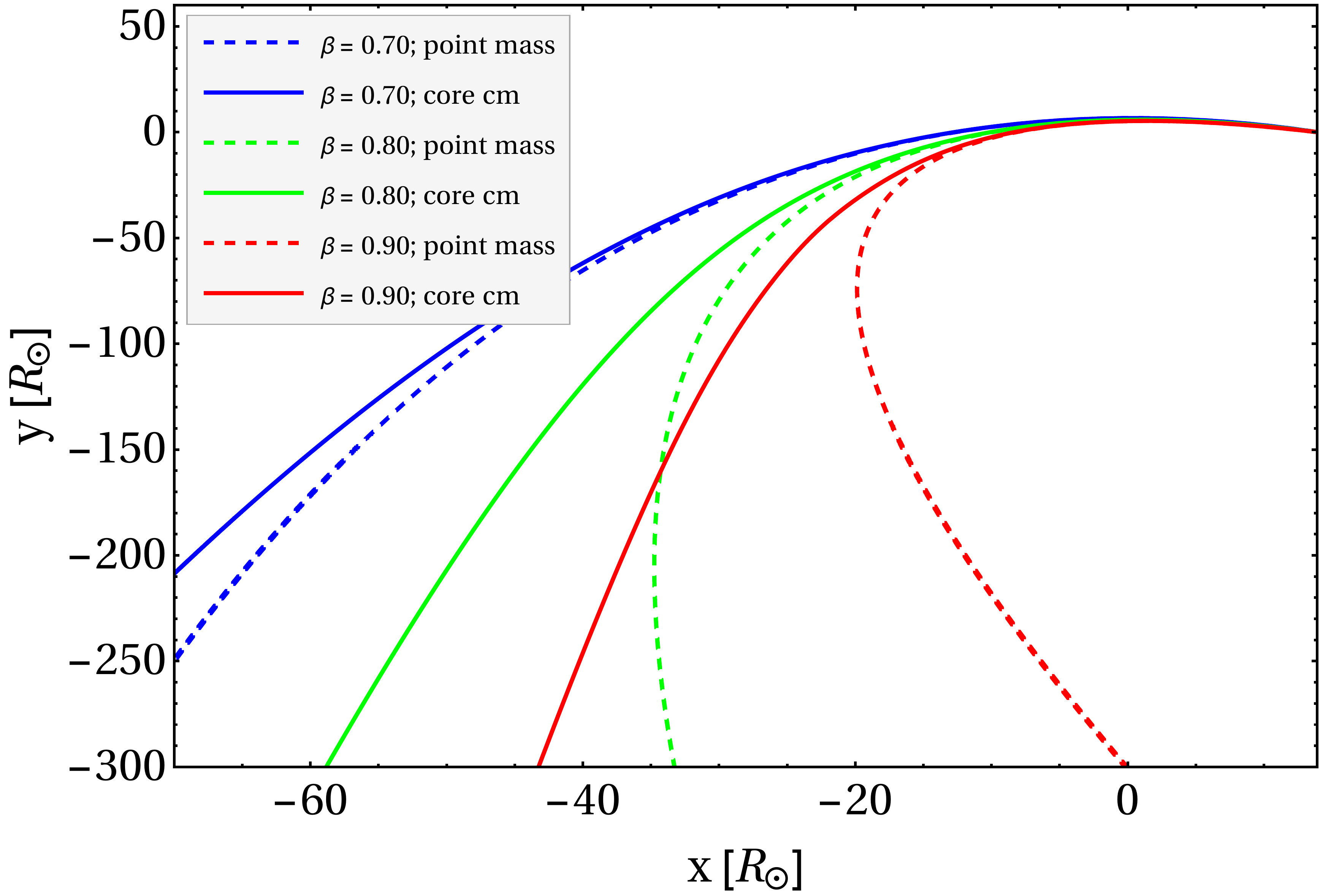}	
	\caption{\textbf{Left panel:} Kick velocity of the core as normalised to the surface escape velocity of the intact star is plotted with time. \textbf{Right panel:} Deviation in remnant core trajectory is shown for three values of $\beta = 0.7, 0.8, 0.9$. Dashed curves indicate the trajectory of the star if it was a point mass.}
	\label{fig.trajec}
\end{figure}

\subsection{Observables: Kick velocity and trajectory deviation}
Change in the specific orbital energy and OAM of the bound core COM leads to the deviation in the geodesic of the core. The whole phenomenon is depicted in Figure \ref{fig.density}. As seen from the column density plots of the stellar partial disruption, the tidal tails are asymmetric, with the inward tail having more mass than the outer tail. After the tidal interaction, the remnant core is recoiled back to its spherical shape and deviates from its initial trajectory. It also becomes a hypervelocity star acquiring kick velocity defined as $v_{\text{kick}}= \sqrt{2(\epsilon-\epsilon_{\text{in}})}$. It should be noted from Figure \ref{fig.eob_time} that the change in specific OAM of the core is $\sim 5\%$ for $\beta=0.9$ while the change in specific orbital energy is only $ \sim 0.00011\%$. Therefore, for this parameter set: $M=100M_\odot, M_\star= 1 M_\odot, R_\star= 1R_\odot$, we find significant deviations in the core trajectory as well as the kick velocity (see Figure \ref{fig.trajec}). The bottom left panel shows the kick velocity of the self-bound core normalised to the surface escape velocity of the star $ v_{\text{esc}}= \sqrt{2GM_\odot/R_\odot} \simeq 617 $ km s$^{-1}$. The bottom right panel shows the deviation in trajectories for three values of $\beta = 0.7, 0.8, 0.9 $. As expected, for higher values of $\beta$, more asymmetry leads to a larger deviation in trajectory and a stronger kick. As already mentioned, peak relative deviation in specific OAM gets lower with $q$, whereas peak relative deviation in specific orbital energy is higher with $q$. An interesting consequence is that significant trajectory deviation can only happen when strong asymmetry is present, i.e., $ M \lesssim \mathcal{O}(10^3 M_\odot)$. However, the star can achieve kick velocity for any black hole mass as far as it is partially disrupted.  

\subsection{Resolution dependence}
Simulation results presented above are also reproduced with a lower resolution with $N=10^5$ particles. It is found that the results are better for lower $\beta$ values for higher resolution. It is reasonable since tidal tails have a lower mass for lower $\beta$, thus poorly resolved for lower $N$. Consequently, specific OAM being highly sensitive to particles for OAM transfer, fails to attain constant value showing small fluctuations in case of lower $\beta$. However, for intermediate $\beta$ values, we find the behavior stable. For $\beta=0.9$, we find $l_{\text{core}}$ drops quicker in case of higher resolution. There are oscillations near the peaks in Figure \ref{fig.eob_time} top left panel and Figure \ref{fig.trajec} left panel. It is argued that the stellar core is compressed and stretched causing shock waves to travel throughout the stellar body. During this time, significant orbital angular momentum and energy transfer between the particles cause their average values to oscillate.

\section{Summary and Discussion}
\label{sec5}
In this paper, we have shown that a star approaching a black hole in a parabolic orbit changes its trajectory after the tidal interaction. We have performed SPH simulations to demonstrate partial tidal disruption due to a Schwarzschild black hole. Our choice of parameters is set to produce asymmetry in tidal force fields around the stellar body, causing more deformation towards the black hole. Consequently, tidal tails are formed having mass difference which increases with $\beta$. Similarly, the stellar core, after tidal stripping, loses mass, and the core mass decreases with $\beta$ as well. Due to the mass asymmetry in the tidal tails, specific orbital energy and orbital angular momentum of the core increase after the tidal encounter, and the increment amount goes higher with $\beta$. Interestingly, we found that the relative deviation in specific orbital energy increases with mass ratio $q$. However, we find the opposite behavior for relative deviation in orbital angular momentum. A consequence of this is that there are less deviations in trajectories with higher mass ratios. On the other hand, kick velocity is larger for higher black hole mass. It leads to the important observation that trajectory deviations are prominent for stellar and IMBHs. Study of stellar orbital dynamics around IMBHs could be an important tool for detecting black holes in this specified range. 

After single or multiple tidal interactions, the stellar orbit is expected to deviate from its point-mass trajectory. There are considerable observational aspects to it that we aim to perform in a subsequent upgradation of the draft. In the case of elliptical bound orbits, multiple asymmetric tidal interactions during multiple pericentre passages may lead to significant deviations in orbital precession and changes in the length of the semi-major axis of the elliptical orbit. As mentioned earlier, extensive studies on the orbital dynamics of S-stars around Sgr A* have revealed stellar properties, black hole parameters, dark matter profile and test gravity theories. The innermost S2 star  ( $\sim 15M_\odot$, $\sim 5R_\odot$) with pericentre $120$ AU is far away from its tidal radius $\sim 1$ AU \cite{Heissel}. However, stellar orbits of similar stars around IMBHs having mass $ \leqslant 10^4 M_\odot$ are expected to undergo tidal trajectory deviations revealing the presence of the IMBH. Such observations of stellar dynamics can be used as a tool to detect IMBHs having mass in globular clusters and infer their properties.  

It is also noted that the nature of the interior of stellar objects, especially stellar compactness, and equation of state directly influences its tidal compressibility. In addition, during the fallback of the tidal debris, observed light curves can lead to understanding the nature of the stellar interior. Thus, tidal kick and trajectory deviations may differ depending on the stellar interior.

It leads us to comment on the observable features of the bound core \cite{Alexander}. The initial star, after its disruption, mostly loses its outer layer. Due to violent tidal distortion, shock heating, and mixing, the convective core size and its mean molecular weight increases. In case of higher mass loss, a core of mass $\sim 0.4 M_\odot$ or less becomes fully convective. As a result, the core becomes more luminous. After the re-accretion of material from the tidal tails, the core becomes redder and ends up on the Hayashi track once again, where it starts contracting on a Kelvin-Helmholtz timescale before nuclear fusion initiates. In case of red giants, hydrogen rich outer envelope, being loosely bound to the core, is easily stripped off due to partial tidal disruption. As a consequence, its inner Helium core is exposed, and the effective temperature goes up, making it bluer yet smaller than a main sequence star with the same mass and age \cite{de Marchi}. As per the argument of Alexander and Livio \cite{Alexander}, a significant number of stars are tidally scattered due to supermassive black holes. They also predicted an abundance of bluer stars than red giants in galactic centres.  

The simulations are performed using true general relativistic tidal forces as compared to other studies that follow Newtonian or pseudo-Newtonian external acceleration. However, for the parameters we used ($M = 100 M_\odot, M_\star = 1 M_\odot, R_\star = 1R_\odot$), the star corresponds to very large pericentre as compared to the Schwarzschild radius, i.e., $r_p \sim 10^4 r_s$. For the same reason, spin of the black hole is neglected. Thus, effectively, the physical phenomena described here are valid in the Newtonian regime as well. For near-horizon disruptions, true relativistic external acceleration will bring significant changes in the tidal interaction, which we aim to perform in a future publication. \\

\noindent
{\bf Acknowledgements}\\

\noindent
We acknowledge the High Performance Computing (HPC) facility at IIT Kanpur, India, where the numerical computations were carried out. 


\end{document}